\documentclass[journal=jpclcd,manuscript=letter,layout=onecolumn]{achemso}
\setkeys{acs}{articletitle = true}

\SectionNumbersOn

\usepackage[version=3]{mhchem} 
\usepackage[T1]{fontenc}
\usepackage{amsmath}
\usepackage{amssymb}
\usepackage{mathrsfs}

\usepackage{listings}

\usepackage{subcaption}
\usepackage{color}
\usepackage{booktabs}
\usepackage{multirow}
\usepackage{units}
\usepackage{bm}
\usepackage{braket}
\usepackage{rotating}
\usepackage{mathtools}

\newcommand{\ReSpect}{\textsc{ReSpect}}

\DeclareMathOperator{\Tr}{Tr}
\renewcommand{\vec}[1]{\boldsymbol{#1}}

\author{Torsha Moitra}
\affiliation{Hylleraas Centre for Quantum Molecular Sciences, Department of Chemistry, UiT The Arctic University of Norway, 9037 Troms{\o}, Norway}
\author{Lukas Konecny}
\affiliation{Hylleraas Centre for Quantum Molecular Sciences, Department of Chemistry, UiT The Arctic University of Norway, 9037 Troms{\o}, Norway}
\alsoaffiliation{Max Planck Institute for the Structure and Dynamics of Matter, Center for Free Electron Laser Science, Luruper Chaussee 149, 22761 Hamburg, Germany}
\author{Marius Kadek}
\affiliation{Hylleraas Centre for Quantum Molecular Sciences, Department of Chemistry, UiT The Arctic University of Norway, 9037 Troms{\o}, Norway}
\alsoaffiliation{Department of Physics, Northeastern University, Boston, Massachusetts 02115, USA}
\alsoaffiliation{Algorithmiq Ltd., Kanavakatu 3C, FI-00160 Helsinki, Finland}
\author{Angel Rubio}
\affiliation{Max Planck Institute for the Structure and Dynamics of Matter, Center for Free Electron Laser Science, Luruper Chaussee 149, 22761 Hamburg, Germany}
\alsoaffiliation{Center for Computational Quantum Physics (CCQ), The Flatiron Institute, 162 Fifth Avenue, New York, New York 10010, USA}
\alsoaffiliation{Nano-Bio Spectroscopy Group, Departamento de F\'{i}sica de Materiales, Universidad del Pa\'{i}s Vasco, 20018 San Sebastian, Spain}
\author{Michal Repisky}
\affiliation{Hylleraas Centre for Quantum Molecular Sciences, Department of Chemistry, UiT The Arctic University of Norway, 9037 Troms{\o}, Norway}
\alsoaffiliation{Department of Physical and Theoretical Chemistry, Faculty of Natural Sciences, Comenius University, 81499 Bratislava, Slovakia}
\email{michal.repisky@uit.no}

\begin{tocentry}
	\includegraphics[]{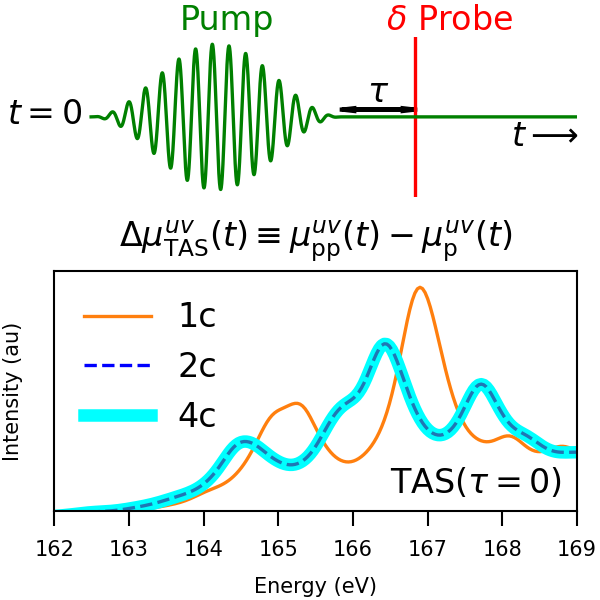}
\end{tocentry}

\title[\texttt{achemso} demonstration]
{Accurate Relativistic Real-Time Time-Dependent Density Functional Theory for Valence and Core Attosecond Transient Absorption Spectroscopy}

\begin{document}

\begin{abstract}
\noindent
First principle theoretical modeling of out-of-equilibrium processes observed in 
attosecond pump--probe transient absorption spectroscopy (TAS) triggering pure electron dynamics 
remains a challenging task, specially for heavy elements and/or core excitations containing fingerprints of scalar and spin-orbit relativistic effects. 
To address this, we formulate a
methodology for simulating TAS within the relativistic real-time
time-dependent density functional theory (RT-TDDFT) framework, for both the
valence and core energy regime.
Especially for TAS, full four-component (4c) RT simulations are feasible but computationally demanding. Therefore, 
in addition to the 4c approach, we also introduce the atomic mean-field exact two-component (amfX2C) Hamiltonian accounting for one- and two-electron picture-change corrections within
RT-TDDFT. amfX2C preserves the accuracy of the parent 4c method at a fraction of its
computational cost. Finally, we apply the methodology to study
valence and near L$_{2,3}$-edge TAS processes of experimentally relevant systems and
provide additional physical insights using relativistic non-equilibrium
response theory. 
\end{abstract}


Transient absorption spectroscopy (TAS) is a powerful non-linear technique for
investigating ultrafast state-resolved processes and electronic superposition
using two pulses, \textit{viz} pump and time-delayed probe
pulse.~\cite{Berera2009,Sommer2016,TAS-HHG-Leone} TAS offers additional degrees
of freedom, namely, pump field features like shape, amplitude, carrier
frequency, direction, and pump-probe time delays, absent in conventional
spectroscopy. Of which, the effect of time-delay between the pulses is most
commonly followed experimentally.~\cite{kling2008attosecond,tzallas2011extreme}
Since long, the emphasis has been on the femto- to pico-second timescales,
which involved time-resolving processes initiated by nuclear
motions.~\cite{chergui-2004-chemrev-pump-probe,Loh2013,Shelby2016,curchod2018-nucleardyn}
However, with the advent of atto-second pulses, there has been an ever
increasing interest in studying sub-femtosecond timescale processes, focusing
on the motion of electrons on their natural time-scale, with minimal influence of
nuclear degrees of
freedom.~\cite{hentschel2001attosecond,baltuvska2003attosecond,kienberger2004atomic,diels2006ultrashort,goulielmakis2010real,ElectronDynamics-2011-Science,leone2014-attosecond,kraus2015measurement,Sommer2016,PhysRevLett.117.093002,nisoli2017attosecond}  
Simultaneously, generation of intense isolated soft X-ray free electron laser
pulses with sub-femtosecond temporal widths has been achieved recently,
promoting investigations  of attosecond resolved spectroscopies involving core
orbitals.~\cite{canova2015optical,duris2020softXray-attopulse} Due to their
characteristic element specificity and unprecedented temporal resolution,
ultrafast X-ray spectroscopies are emerging as indispensable tools for
biological and material sciences.~\cite{Chen2013-XTAS,Leone2018-XTAS}
Especially X-ray transient absorption spectroscopy (XTAS) which employs short
X-ray probe pulses is now extensively being used to investigate electron
dynamics of molecules and solid-state
systems.~\cite{cavalieri2007attosecond,Chen2013-XTAS,Loh2013,schultze2014attosecond,lucchini2016attosecond,Leone2018-XTAS,picon2019attosecond}
This dictates the need to develop reliable theoretical tools to aid the
interpretation and prediction of such phenomena.

With regards to simulating the response of the laser pulses relevant to pump-probe
spectroscopies, real-time (RT) formalism offers a straightforward
approach~\cite{yabana1996rt,tsolakidis2002rt,isborn2008rt,Lopata2011,Lopata-JCTC-8-3284,sato2013time,DeGiovannini2013,Lian2018,schelter2018accurate,pedersen2019symplectic,pedersen2020interpretation,vila2020real,Lopata2020-XTAS,PhysRevA.102.023115,PhysRevA.105.023103,schreder2021local,mattiat2021recent,Liu2022,Lopata2022-TAS}
over response theory based methods as the former is applicable for a large
range of field intensities, and resembles the experimental setup in a natural
way. Addressing core level spectroscopies has added complexities, as it requires the
inclusion of scalar (SC) and spin-orbit (SO) relativistic effects, which are
most reliably described by multi-component relativistic quantum chemical
methods.  Here, the ``gold standard'' is the four-component (4c) methodology
including both SC and SO effects variationally via the one-electron Dirac
Hamiltonian in combination with instantaneous Coulomb interactions among the
particles. Since the 4c treatment can be time consuming, especially for RT
simulations, researchers have focused on the development of approximate 2c
Hamiltonians.~\cite{Saue2011} An approach that has gained wide popularity in
recent years is the exact two-component (X2C) Hamiltonian as 
it reduces the original 4c problem by half, requiring only a few simple algebraic
manipulations.~\cite{Heully1986,Jensen2005,Kutzelnigg2005,Liu2007,Ilias2007,Liu2009}
There exist several variants of X2C Hamiltonian, ranging from a pure
one-electron X2C (1eX2C) Hamiltonian where two-electron (2e) interactions are entirely
omitted from the X2C decoupling transformation,~\cite{Saue2011,Konecny2016} to
a molecular mean-field X2C (mmfX2C) Hamiltonian where the decoupling is performed
\emph{after} 4c molecular self-consistent field (SCF) calculations.~\cite{Sikkema2009}
Inbetween there are several X2C Hamiltonian models that extend 1eX2C by
including 2e interactions approximately via (i) element and angular-momentum
specific screening factors in the evaluation of one-electron SO
integrals~\cite{Blume1962,Blume1963}; (ii) a mean-field SO
approach~\cite{Hess1996} which has been the basis for the widely popular AMFI
module~\cite{AMFI}; and (iii) an approach that exploits atomic model densities
obtained within the framework of Kohn--Sham DFT.~\cite{Wullen2005,Peng2007} The
screening factors of type (i) are sometimes referred to as ``Boettger factors''
or as the (modified) screened--nuclear--spin--orbit approach
((m)SNSO)~\cite{Boettger2000,Filatov2013}. Recently, an atomic mean-field
(amfX2C) as well as an extended atomic mean-field (eamfX2C) approach have been
presented within the X2C framework~\cite{amfx2c}, extending some of earlier
ideas of Liu and Cheng~\cite{LiuCheng2018} by comprising the full 2e SO and SC
corrections, regardless whether they arise from the Coulomb, Coulomb--Gaunt, or
Coulomb--Breit Hamiltonian. Moreover, this ansatz takes into account the
characteristics of the underlying correlation framework, \emph{viz.},
wave-function theory or (KS-)DFT, which enables tailor-made
exchange--correlation (xc) corrections to be introduced.~\cite{amfx2c}
While all above mentioned relativistic methods lie within the static
time-independent regime, extensions to the RT framework~\cite{li2020real}
were recently formulated both at 4c~\cite{Repisky2015,Kadek2015,de2020pyberthart} 
and 2c~\cite{Konecny2016,goings2016real,koulias2019relativistic} level.

In this letter, we present a novel theoretical methodology to address several
fascinating characteristics associated with TAS and its simulation.  First, to
understand the physics governing TAS, we generalize the non-equilibrium
response theory formalism to incorporate complex orbitals necessary for
relativistic theory. This facilitates us to interpret unique spectral
observations inherent to TAS. Next, to provide first-principle computational
approach for simulation of TAS across the entire Periodic Table and/or core
atomic region(s), we implement a relativistic variant of RT time-dependent
density functional theory (TDDFT) based on 4c Dirac--Coulomb Hamiltonian. This
allows us to account for SC and SO relativistic effects variationally, thereby
significantly broadening the applicability of the tool. Since the gold standard
4c method is still computationally demanding and popular one-electron 1eX2C is
not sufficiently accurate in comparison to its 4c reference,~\cite{Konecny2023}
finally we formulate and implement for the first time a simple yet numerically
accurate amfX2C Hamiltonian in context of RT framework.  It incorporates all
spin-free and spin-dependent relativistic picture-change corrections
originating from X2C transformation, giving remarkable agreement with reference
4c results. Consequently, we apply the amfX2C Hamiltonian to provide physical
insights into TAS near valence and L$_{2,3}$-edge for experimentally relevant
systems. 


In order to lay the theoretical foundations for describing pump-probe spectroscopy,
we consider in this work the non-overlapping regime, in which the probe field/pulse 
$\bm{\mathcal{F}}(t)$ is applied at or after the end of the pump pulse 
$\bm{\mathcal{E}}(t)$. The pump pulse takes the form 
\begin{equation}
	\bm{\mathcal{E}}(t)
	=
  \bm{n} \mathcal{E}(t)
  =
	\bm{n} \mathcal{E}_0 
	\cos^2 \left(\pi \frac{t-t_0}{T} \right)
	\sin(\omega_0 t + \phi),
	\label{eq:pump}
\end{equation}
and is characterized by the amplitude $\mathcal{E}_0$, 
shape ($\cos^2$-enveloped $\sin$ function), carrier frequency $\omega_0$, 
time duration $T$, carrier--envelope phase (CEP) $\phi$, and polarization along
the unit vector $\bm{n}$.
The pump pulse is centered at $t_0$, and is zero (inactive) outside of the
window of size $T$, \emph{i.e.} for $t < t_0 - \tfrac{T}{2}$ and $t > t_0+
\tfrac{T}{2}$.
In practice, the duration of the pump is chosen to be an integer number of
carrier periods and the time $t=0$ corresponds to the onset of the pump, i.e.
$t_0= \tfrac{T}{2}$, and the CEP is set to 0. 
For the probe pulse, we use the analytical form of a delta field~\cite{Repisky2015}
\begin{equation}
	\bm{\mathcal{F}}(t)
	=
	\bm{m} \mathcal{F}(t)
	=
	\bm{m} \mathcal{F}_0 \delta(t-(T+\tau)),
	\label{eq:probe}
\end{equation}
characterized by its amplitude $\mathcal{F}_0$, polarization direction $\bm{m}$,
and positioned at $T+\tau$, where $\tau$ is a time-delay between pump and probe pulse.

To get physical insight into pump--probe processes, it is instructive to
examine them from the point of non-equilibrium response
theory.~\cite{stefanucci-response} First, we derive the non-perturbative form
of the response to the probe field applied to the non-stationary
(out-of-equilibrium) state $\ket{\Psi[\bm{\mathcal{E}}]}$ generated by the
previous pump pulse lasting from time 0 to $T$.  The time evolution of a system
after the end of the pump is described by the Hamiltonian,
\begin{equation}
  \label{eq:Hamiltonian}
	\hat{H}(t) = \hat{H}_{0} - \sum_{u\in x,y,z}\mathcal{F}_{u}(t)\hat{P}_u
	;\qquad
	\hat{P}_u = - \sum_{i=1}^{N} \hat{r}_{i,u},
\end{equation}
where $\hat{H}_{0}$ is the static Hamiltonian describing the $N$ electron
system itself, and $\sum_{u\in x,y,z}\mathcal{F}_{u}(t)\hat{P}_u$ denotes the
coupling of the molecular system to the probe pulse via the electric dipole
operator $\bm{\hat{P}}$. Note that we have assumed the dipole approximation and
we work in the length gauge all over the paper.  The wave function
$\ket{\Psi(t)}$ for times $t>T+\tau$ can be expressed using unitary evolution
operators~\cite{Tannor} that propagate the state $\ket{\Psi[\bm{\mathcal{E}}]}$
from the time $t=T$. To isolate the singularity caused by the $\delta$-type
probe pulse, we split the time propagation into three parts using a sequence of
evolution operators. First, we propagate from the time $t=T$ to $t=T+\tau$.
Here, the evolution operator has the simple form of $e^{-i\hat{H}_0\tau}$
(where $T$ cancelled out), since both the pump and probe pulses are not active
in this interval, and the time propagation is determined only by the static
Hamiltonian $\hat{H}_0$.  Second, we handle the probe perturbation by
propagating the wave function on the infinitesimal time interval from
$t=T+\tau-\epsilon$ to $t=T+\tau+\epsilon$ as $\epsilon\rightarrow 0$. Such a
``step'' propagation can be expressed in the closed form~\cite{Repisky2015} as
$e^{-i\hat{Q}_v}$, where $\hat{Q}_v \equiv - \mathcal{F}_0\hat{P}_v$. Third,
the evolution from $t=T+\tau$ to an arbitrary $t$ is, like in the first case,
determined by the static Hamiltonian as $e^{-i\hat{H}_0(t-T-\tau)}$. Hence, the
wave function is given by
\begin{equation}
  \label{eq:evolWaveFunc}
  \ket{\Psi(t)} = e^{-i\hat{H}_0(t-T-\tau)} e^{-i\hat{Q}_v} e^{-i\hat{H}_0\tau} \ket{\Psi[\bm{\mathcal{E}}]}.
\end{equation}
In the basis of stationary eigenstates $\ket{\Phi_j}$ of the static
Hamiltonian $\hat{H}_0$, the state $\ket{\Psi[\bm{\mathcal{E}}]}$ is expressed as
a linear combination  
\begin{align}
	\label{eq:wavepacket}
	\ket{\Psi[\bm{\mathcal{E}}]} = \sum_j c_j(\bm{\mathcal{E}}) \ket{\Phi_j}
  ;\qquad
  c_j(\bm{\mathcal{E}}) = \braket{\Phi_j|\Psi[\bm{\mathcal{E}}]}.
\end{align}
Using the resolution-of-the-identity $\sum_j\ket{\Phi_j}\bra{\Phi_j} =
\hat{1}$, the non-perturbative electric dipole response to the probe pulse
$P_u[\bm{\mathcal{E}},\bm{\mathcal{F}}](t) =
\braket{\Psi(t)|\hat{P}_u|\Psi(t)}$ can be written in the frequency domain as
\begin{equation}
  \label{eq:nonpertResponse}
  P_u[\bm{\mathcal{E}},\bm{\mathcal{F}}](\omega) = i\sum_{jkmn} c^*_j(\bm{\mathcal{E}}) c_n(\bm{\mathcal{E}}) e^{i\omega_{jn}\tau} \frac{\left(e^{iQ_v}\right)_{jk}P_{u,km}\left(e^{-iQ_v}\right)_{mn}}{\omega_{km}+\omega+i\Gamma},
\end{equation}
where $\omega_{jk} \equiv \epsilon_j - \epsilon_k$ is the difference between
the energies of the stationary states,
for an operator $\hat{A}$, $A_{jk} \equiv \braket{\Phi_j|\hat{A}|\Phi_k}$
is its matrix element, $P_u(\omega) = \int_{T+\tau}^{\infty}
P_u(t)e^{(i\omega-\Gamma)(t-T-\tau)} dt$ and a damping parameter $\Gamma$ is
introduced to regularize the Fourier integral over oscillating functions and
facilitate its practical evaluation in simulations with a finite time length.
Using the expansion $e^{-iQ_v} \equiv e^{i\mathcal{F}_0 P_v} = \mathbb{I} +
i\mathcal{F}_0 P_v + \ldots$ in Eq.~\eqref{eq:nonpertResponse} allows us to
understand the most dominant effects on the resulting spectra, \emph{i.e.}
those that appear even in the weak probe field limit. After rearranging the
summation indices, we obtain the first terms
\begin{equation}
  \label{eq:response0}
  P^{(0)}_u[\bm{\mathcal{E}},\bm{\mathcal{F}}](\omega) = i\sum_{jk} c_j(\bm{\mathcal{E}})c^*_k(\bm{\mathcal{E}}) e^{-i\omega_{jk}\tau} \frac{P_{u,kj}}{\omega-\omega_{jk}+i\Gamma},
\end{equation}
and
\begin{equation}
  \label{eq:response1}
  \frac{P^{(1)}_u[\bm{\mathcal{E}},\bm{\mathcal{F}}](\omega)}{\mathcal{F}_0} = \sum_{jkm} c_j(\bm{\mathcal{E}})c^*_k(\bm{\mathcal{E}}) e^{-i\omega_{jk}\tau} \frac{P_{v,km}P_{u,mj}}{\omega-\omega_{jm}+i\Gamma} + \text{c.c.}(\omega\rightarrow -\omega),
\end{equation}
where c.c.$(\omega\rightarrow -\omega)$ labels the complex conjugation while
flipping the sign of the frequency. We note, that in the relativistic theory
presented here, the stationary wave functions are complex in general. As a
consequence, the matrix elements $P_{v,km}$ can no longer be assumed to be
real.  Expressions in Eqs.~\eqref{eq:response0} and~\eqref{eq:response1} are in
contrast with the equilibrium response theory where a response function is
defined in terms of a specific eigenstate of $\hat{H}_0$, \emph{e.g.}
$\ket{\Psi[\bm{\mathcal{E}}]} \equiv \ket{\Phi_0}$, and $c_j(\bm{\mathcal{E}})
\equiv \delta_{j0}$ --- assuming this would yield the equilibrium response
functions.

The process of electronic absorption, that is of interest here, is associated
with the imaginary part of the electric dipole--electric dipole response
function in the frequency domain ($\chi_{\hat{P}_u,\hat{P}_v}[\bm{\mathcal{E}}]
(\omega,T+\tau) \equiv
P^{(1)}_u[\bm{\mathcal{E}},\bm{\mathcal{F}}](\omega)/\mathcal{F}_0$).~\cite{NormanRuudSaue}
The resulting expression for the response function was presented by other
authors~\cite{Walkenhorst2016}, while assuming that the eigenstates
$\ket{\Phi_j}$, and hence the matrix elements $P_{v,km}$, are real-valued.
However, this is not the case in the relativistic theory, and we here proceed
by deriving the relativistic extension for the response function without making
this assumption. Let $\phi_j$ denote the phase of the complex amplitudes $c_j$,
\emph{i.e.} $c_j = |c_j|e^{i\phi_j}$. By defining $\Theta_{jk}(\tau) \equiv
\phi_j - \phi_k - \omega_{jk}\tau$ and using $1/(\omega + i\Gamma) =
\mathcal{R}(\omega) - i \mathcal{L}(\omega)$, where $\mathcal{L}$ and
$\mathcal{R}$ represent Lorentzian and Rayleigh lineshape functions,
respectively, we can rearrange the terms in Eq.~\eqref{eq:response1} and
extract the imaginary part as
\begin{align}
  \Im\chi_{\hat{P}_u,\hat{P}_v}[\bm{\mathcal{E}}] (\omega,T+\tau) &= \sum_{jkm} |c_j(\bm{\mathcal{E}})c_k(\bm{\mathcal{E}})| \Re\left(P_{v,km}P_{u,mj}\right) \nonumber\\
  &\times \big[\sin\Theta_{jk}(\tau) \mathcal{R}(\omega-\omega_{jm}) - \cos\Theta_{jk}(\tau) \mathcal{L}(\omega-\omega_{jm}) \nonumber\\
  &+ \sin\Theta_{jk}(\tau) \mathcal{R}(\omega+\omega_{jm}) + \cos\Theta_{jk}(\tau) \mathcal{L}(\omega+\omega_{jm})\big] \nonumber\\
  &+ \sum_{jkm} |c_j(\bm{\mathcal{E}})c_k(\bm{\mathcal{E}})| \Im\left(P_{v,km}P_{u,mj}\right) \nonumber\\
  &\times \big[ \cos\Theta_{jk}(\tau) \mathcal{R}(\omega-\omega_{jm}) + \sin\Theta_{jk}(\tau) \mathcal{L}(\omega-\omega_{jm}) \nonumber\\
  &+ \cos\Theta_{jk}(\tau) \mathcal{R}(\omega+\omega_{jm}) - \sin\Theta_{jk}(\tau) \mathcal{L}(\omega+\omega_{jm})\big].
  \label{eq:response1Imag}
\end{align}
The first part of this expression proportional to the real part
$\Re\left(P_{v,km}P_{u,mj}\right)$ coincides with the result of Walkenhorst
\emph{et al.}~\cite{Walkenhorst2016} for $P_{v,km}\in\mathbb{R}$.  The second
part is nonzero only for theories that lead to complex stationary states
(\emph{e.g.} when the Hamiltonian $\hat{H}_{0}$ is not real-valued), such as
the relativistic theory with SO effects included. We note here, that this
distinction between the theories based on complex and real orbitals is a unique
feature of the non-equilibrium response function, and the differences between
the formulations vanish when the non-stationary state
$\ket{\Psi[\bm{\mathcal{E}}]}$ is replaced with a single eigenstate (for
spatially isotropic values $\Im\chi_{\hat{P}_u,\hat{P}_u}$).  Finally, we
mention that the diagonal terms $j=k$ in the sums in
Eq.~\eqref{eq:response1Imag} can be isolated to study separately the time
delay-independent contribution and interference term that depends on the time
delay $\tau$. The interference term is a signature of the non-stationary state,
leading to the spectral dependence on $\tau$. $\Theta_{jk}$ combines the real
and imaginary part of the response function and interchanges the $\mathcal{L}$
to $\mathcal{R}$ lineshapes, and vice-versa. The imprints of
Eq.~\eqref{eq:response1Imag} can be directly connected to the simulated
spectral observations and are discussed later.~\cite{Walkenhorst2016,stefanucci-response}

Instead of using the non-equilibrium response theory to obtain TAS spectra, we
work directly in time domain and evolve a molecular system of interest by the
Liouville--von Neumann equation-of-motion (EOM). For Kohn--Sham TDDFT in an
orthonormal basis, EOM takes the form
\begin{equation}
\label{eq:LvN}
i \frac{\partial \mathbf{D}(t)}{\partial t} = [\mathbf{F}(t),\mathbf{D}(t)]
,
\end{equation}
where $\mathbf{D}(t)$ is the time-dependent reduced one-electron density matrix
describing the state of the system at time $t$ and $\mathbf{F}(t)$ is the Fock
matrix driving the time evolution and characterizing the molecular system
itself as well as its interaction with the external pump--probe fields. In
practice, Eq.~\eqref{eq:LvN} is solved numerically by propagating
$\mathbf{D}(t)$ in time as detailed in Ref.~\citenum{Lopata2011,Repisky2015},~\citenum{castro2004propagators} and~\citenum{gomez2018propagators} as
well as in Section S1. 

The theoretical modelling of core--level spectroscopies is a challenging task
because the spectra feature a fine structure due to scalar (SC) and spin--orbit
(SO) relativistic
effects~\cite{Kadek2015,Norman2018,Kasper2020,Bokarev2020,Besley2021,Konecny2022}.
With this in mind, we have extended our recent probe-only four-component (4c)
RT-TDDFT implementation~\cite{Repisky2015,Kadek2015,ReSpect} to the realm of
pump--probe experiments. Assuming an orthonormalized atomic orbital (AO) basis, 
the 4c Fock operator suitable for TAS has the
following matrix form
\begin{align}
  \label{eq:4cFock}
  F^{\mathrm{4c}}_{\mu\nu}(t)
  =
  F^{\mathrm{4c}}_{\mu\nu}[\boldsymbol{\mathcal{E}},\boldsymbol{\mathcal{F}}](t)
  & =
	h^{\mathrm{4c}}_{\mu\nu} 
  -
  \sum_{u\in x,y,z}
  P^{\mathrm{4c}}_{u,\mu\nu}
  \mathcal{E}_u(t)
  -
  \sum_{u\in x,y,z}
  P^{\mathrm{4c}}_{u,\mu\nu}
  \mathcal{F}_u(t)
  \nonumber
  \\
  & +
  \sum_{\kappa\lambda}
  G^{\text{4c}}_{\mu\nu,\kappa\lambda}
  D^{\text{4c}}_{\lambda\kappa}(t,\boldsymbol{\mathcal{E}},\boldsymbol{\mathcal{F}})
  +
  \sum_{u\in0-3}
  \int v^{xc}_{u}\!\left[\boldsymbol{\rho}^\text{4c}(\vec{r},t,\boldsymbol{\mathcal{E}},\boldsymbol{\mathcal{F}})\right]
  \Omega_{u,\mu\nu}^{\text{4c}}(\vec{r})\,d^{3}\vec{r}
  ,
\end{align}
where terms on the right hand side include the one-electron Dirac operator
($\mathbf{h}^{\mathrm{4c}}$), particle-field interactions via the electric
dipole moment operator ($\mathbf{P}^{\mathrm{4c}}_{u}$), two-electron (2e)
Coulomb and exchange interactions, and the exchange--correlation (xc)
contribution. Here, the 2e term involves the matrix of generalized
anti-symmetrized electron repulsion integrals
\begin{equation} 
   \label{eq:eri}
   G^{\text{4c}}_{\mu\nu,\kappa\lambda}
    =
   \mathcal{I}^{\text{4c}}_{\mu\nu,\kappa\lambda}
   -
   \zeta\mathcal{I}^{\text{4c}}_{\mu\lambda,\kappa\nu}
   ;\qquad
   \mathcal{I}^{\text{4c}}_{\mu\nu,\kappa\lambda}
   \coloneqq
   \iint
   \Omega_{0,\mu\nu}^{\text{4c}}(\vec{r}_{1})
   r_{12}^{-1}
   \Omega_{0,\kappa\lambda}^{\text{4c}}(\vec{r}_{2})
   d^{3}\vec{r}_{1}d^{3}\vec{r}_{2}
   ,
\end{equation}
with a scalar scaling factor $\zeta$ for exchange interaction, whereas the xc term
requires a noncollinear xc potential $v_u^{xc}$, given within a
generalized gradient approximation (GGA) by derivatives of the nonrelativistic
xc energy density ($\varepsilon^{xc}$) with respect to the 4c electronic charge
density ($\rho_{0}^{\text{4c}}$), spin densities ($\rho_{1-3}^{\text{4c}}$), 
as well as their gradients:
\begin{equation}
   \label{eq:vxc:4c}
   v^{xc}_u\!\left[\boldsymbol{\rho}^{\text{4c}}\right]
   =
   \frac{\partial\varepsilon^{xc}}{\partial \rho_u^{\text{4c}}}
   -
   \left( \boldsymbol{\nabla} \cdot
          \frac{\partial\varepsilon^{xc}}{\partial \boldsymbol{\nabla}\rho_u^{\text{4c}}}\right)
   ,\qquad
   \rho_u^{\text{4c}} \coloneqq \rho_u^{\text{4c}}(\vec{r},t,\boldsymbol{\mathcal{E}},\boldsymbol{\mathcal{F}})
   =
   \sum_{\mu\nu}
   \Omega_{u,\mu\nu}^{\text{4c}}(\vec{r}) D^{\text{4c}}_{\nu\mu}(t,\boldsymbol{\mathcal{E}},\boldsymbol{\mathcal{F}})
   .
\end{equation}
More details about our noncollinear extension of nonrelativistic xc
potentials are available in Ref.~\citenum{Komorovsky2019}. In
Eqs.~\eqref{eq:eri} and \eqref{eq:vxc:4c}, $\boldsymbol{\Omega}_{u}^{\text{4c}}$ stands for
the matrix of overlap distribution functions~\cite{ReSpect}
\begin{equation}
   \label{eq:omega0}
   \Omega^{\text{4c}}_{u,\mu\nu}(\vec{r})
   =
   \boldsymbol{X}_{\mu}^{\dagger}(\vec{r}) \boldsymbol{\Sigma}_u \boldsymbol{X}_{\nu}(\vec{r})
   ,
   \qquad
   \boldsymbol{\Sigma}_{u} =
   \begin{pmatrix}
       \boldsymbol{\sigma}_{u} & \mathbf{0}_2 \\
       \mathbf{0}_2  & \boldsymbol{\sigma}_{u}
   \end{pmatrix}
   ,
\end{equation}
defined as the product of \emph{orthonormal} 4c AO basis functions
$\boldsymbol{X}_{\mu}(\bm{r}) \coloneqq \boldsymbol{X}^{\text{RKB}}_{\mu}(\bm{r})$
fulfilling a restricted kinetic balance (RKB) condition in their 
small-component,~\cite{Stanton1984} and 4c operators associated with the electronic 
charge density ($\boldsymbol{\Sigma_0}$) and spin densities ($\boldsymbol{\Sigma}_{1-3}$). 
The latter ones are defined via the $2\times2$ zero matrix ($\mathrm{\mathbf{0}}_{2}$), identity matrix
($\boldsymbol{\sigma}_{0}$), and Pauli spin matrices 
($\boldsymbol{\sigma}_1,\boldsymbol{\sigma}_2,\boldsymbol{\sigma}_3$). 
Note that prior to their use in Eq.~\eqref{eq:LvN}, both Fock and density
matrix are transformed into the basis of ground-state molecular orbitals,
obtained from the solution of self-consistent field equations. 

While simulation of XAS by means of full 4c RT-TDDFT method is nowadays
feasible but computationally challenging~\cite{Kadek2015},
its further extension towards pump--probe experiments
poses additional computational burden. 
Therefore, there is interest in developing
approximate relativistic methods enabling RT calculations in two-component (2c)
regime while maintaining the accuracy of the parent 4c approach. Hence,
alongside of the 4c method that serves as a gold-standard reference, we 
put-forth a first formulation and implementation of
 an atomic
mean-field exact two-component Hamiltonian (amfX2C)
within the realm of
 RT-TDDFT and apply it to simulate (X)TAS spectra. 
 As proven for self-consistent field calculations,
  the simple yet numerically accurate amfX2C approach
accounts for so-called SC and
SO two-electron (2e) and exchange--correlation (xc) picture-change (PC) effects that
arise from the X2C transformation,
in contrast to the commonly used
one-electron X2C (1eX2C) Hamiltonian~\cite{amfx2c}. 
Furthermore,
as demonstrated on simple XAS
spectra of transition metal and actinide compounds, the absence of these PC effects 
in 1eX2C results in a substantial overestimation of L- and M-edge SO
splittings, whereas the amfX2C Hamiltonian reproduces all essential spectral
features such as shape, position and SO splitting in excellent agreement with
4c references, while offering more than 7-fold speed-up.~\cite{Konecny2023} 
A similar acceleration was reported previously 
on RT-TDDFT based on 1eX2C~\cite{Konecny2016}.

The central idea of our RT-TDDFT amfX2C approach is the matrix
transformation of the original 4c EOM in Eq.~\eqref{eq:LvN} to its
diagonally-dominant form using a static (time-independent) unitary
matrix~$\mathbf{U}$. 
By maintaining only the large-component--large-component (LL) block of the
transformed 4c EOM, one arrives at the 2c EOM~\cite{Konecny2016,Konecny2023}
\begin{gather}
  \label{x2c:eom}
  i\frac{\partial \mathbf{\tilde{D}^{\text{2c}}}(t)}{\partial t} 
  = [\mathbf{\tilde{F}}^{\text{2c}}(t),\mathbf{\tilde{D}}^{\text{2c}}(t)]
\end{gather}
with so-called 2c PC transformed Fock and density matrix
\begin{align}
   \tilde{F}^{\mathrm{2c}}_{\mu\nu}(t)
   =
   \Big[
       \mathbf{U}^{\dagger} \mathbf{F}^{\mathrm{4c}}(t) \mathbf{U}
   \Big]^{\mathrm{LL}}_{\mu\nu}
   ,\qquad
   \tilde{D}^{\mathrm{2c}}_{\mu\nu}(t)
   =
   \Big[
       \mathbf{U}^{\dagger} \mathbf{D}^{\mathrm{4c}}(t) \mathbf{U}
   \Big]^{\mathrm{LL}}_{\mu\nu}
   .
\end{align}
In consistency with our previous works we use notation with tildes to
indicate picture-change transformed quantities~\cite{amfx2c,Konecny2023}. Of
significant importance is the observation that the \emph{correctly} transformed
2c Fock matrix involves not only the PC transformed density matrix
($\tilde{\mathbf{D}}^{\mathrm{2c}}$), but also overlap distribution matrix, as
well as one- and two-electron integrals~\cite{amfx2c,Konecny2023}. In connection 
to Eq.~\eqref{eq:4cFock}, our 2c RT-TDDFT (X)TAS reads
\begin{align}
  \label{eq:x2cFock}
  \tilde{F}^{\mathrm{2c}}_{\mu\nu}(t)
  =
  \tilde{F}^{\mathrm{2c}}_{\mu\nu}[\boldsymbol{\mathcal{E}},\boldsymbol{\mathcal{F}}](t)
  & =
	\tilde{h}^{\mathrm{2c}}_{\mu\nu} 
  -
  \sum_{u\in x,y,z}
  \tilde{P}^{\mathrm{2c}}_{u,\mu\nu}
  \mathcal{E}_u(t)
  -
  \sum_{u\in x,y,z}
  \tilde{P}^{\mathrm{2c}}_{u,\mu\nu}
  \mathcal{F}_u(t)
  \nonumber
  \\
  & +
  \sum_{\kappa\lambda}
  \tilde{G}^{\text{2c}}_{\mu\nu,\kappa\lambda}
  \tilde{D}^{\text{2c}}_{\lambda\kappa}(t,\boldsymbol{\mathcal{E}},\boldsymbol{\mathcal{F}})
  +
  \sum_{u\in0-3}
  \int v^{xc}_{u}\!\left[\boldsymbol{\tilde{\rho}}^\text{2c}(\vec{r},t,\boldsymbol{\mathcal{E}},\boldsymbol{\mathcal{F}})\right]
  \tilde{\Omega}_{u,\mu\nu}^{\text{2c}}(\vec{r})\,d^{3}\vec{r}
  .
\end{align}
Note however, that the presence of picture-change transformed overlap distribution matrix
($\boldsymbol{\tilde{\Omega}}^{\text{2c}}$) in both 2e and xc interaction terms makes the
evaluation of $\mathbf{\tilde{F}}^{\mathrm{2c}}$ computationally more demanding
than the original 4c Fock matrix. Therefore, in line with the amfX2C approach 
introduced originally for the static (time-independent) SCF
case~\cite{amfx2c} and extended later to the response theory formalism
involving electric field(s)~\cite{Konecny2023},
$\mathbf{\tilde{F}}^{\mathrm{2c}}$ in Eq.~\eqref{eq:x2cFock} can be approximated 
by a computationally efficient form built with untransformed
(without the tilde) overlap distributions ($\boldsymbol{\Omega}^{\text{2c}}$),
i.e.,
\begin{align}
  \label{eq:amfX2CFock}
  \tilde{F}^{\mathrm{2c}}_{\mu\nu}(t)
  \approx
  F^{\mathrm{amfX2C}}_{\mu\nu}[\boldsymbol{\mathcal{E}},\boldsymbol{\mathcal{F}}](t)
  & =
	\tilde{h}^{\mathrm{2c}}_{\mu\nu} 
  -
  \sum_{u\in x,y,z}
  \tilde{P}^{\mathrm{2c}}_{u,\mu\nu}
  \mathcal{E}_u(t)
  -
  \sum_{u\in x,y,z}
  \tilde{P}^{\mathrm{2c}}_{u,\mu\nu}
  \mathcal{F}_u(t)
  +
  \Delta{\tilde{F}}^{\text{amfX2C}}_{\bigoplus,\mu\nu}
  \nonumber
  \\
  & +
  \sum_{\kappa\lambda}
  G^{\text{2c}}_{\mu\nu,\kappa\lambda}
  \tilde{D}^{\text{2c}}_{\lambda\kappa}(t,\boldsymbol{\mathcal{E}},\boldsymbol{\mathcal{F}})
  +
  \sum_{u\in0-3}
  \int v^{xc}_{u}\!\left[\boldsymbol{\rho}^\text{2c}(\vec{r},t,\boldsymbol{\mathcal{E}},\boldsymbol{\mathcal{F}})\right]
  \Omega_{u,\mu\nu}^{\text{2c}}(\vec{r})\,d^{3}\vec{r}
  .
\end{align}
In fact, Eq.~\eqref{eq:amfX2CFock} defines our 2c amfX2C Fock matrix applied in
actual RT (X)TAS simulations. Here, the picture-change corrections, defined as the
difference between transformed and untransformed 2e and xc interaction terms,
are take into account approximately via $\Delta{\mathbf{\tilde{F}}}^{\text{amfX2C}}_{\bigoplus}$ 
and given by a superposition of corresponding \emph{static} atomic quantities:~\cite{amfx2c,Konecny2023}
\begin{equation} 
   \label{eq:DeltaF2}
   \Delta{\mathbf{\tilde{F}}}^{\text{amfX2C}}_{\bigoplus} 
   =
   \bigoplus_{K=1}^{M}
   \Delta\mathbf{\tilde{F}}^{\text{2c,2e}}_{K}
   +
   \Delta\mathbf{\tilde{F}}^{\text{2c,xc}}_{K}
   ,
\end{equation}
where
\begin{align}
   \Delta\mathbf{\tilde{F}}^{\text{2c,2e}}_{K}
   & = 
   \Big(
   \mathbf{\tilde{G}}^{\text{2c}}_{K}
   -
   \mathbf{G}^{\text{2c}}_{K}
   \Big)
   \mathbf{\tilde{D}}^{\text{2c}}_{K}
   \\
   \Delta\mathbf{\tilde{F}}^{\text{2c,xc}}_{K}
   & = 
   \sum_{u\in0-3}
   \int v_u^{xc}\!\left[\boldsymbol{\tilde{\rho}}^\text{2c}_{K}(\vec{r})\right]
   \boldsymbol{\tilde{\Omega}}_{u}^{\text{2c}}(\vec{r})\,d^{3}\vec{r}
   -
   \sum_{u\in0-3}
   \int v_u^{xc}\!\left[\boldsymbol{\rho}^\text{2c}(\vec{r})\right]
   \boldsymbol{\Omega}_{u}^{\text{2c}}(\vec{r})\,d^{3}\vec{r}
   .
\end{align}
Here, $K$ runs over all atoms in an $M$-atomic system and labels atomic
quantities obtained from independent atomic SCF calculations, each performed in
the orthonormal AO basis of $K$th atom. While a theoretical justification enabling
us to build $\Delta{\mathbf{\tilde{F}}}^{\text{amfX2C}}_{\bigoplus}$ from
static (time-independent) quantities is available in
Ref.~\citenum{Konecny2023}, a pseudo-code highlighting the essential steps for
evaluating  $\Delta{\mathbf{\tilde{F}}}^{\text{amfX2C}}_{\bigoplus}$ is
presented in Ref.~\citenum{amfx2c}.
The final TAS spectra are obtained within the RT-TDDFT framework from the differential induced dipole moment
\begin{equation}
  \label{eq:deltamu_TAS}
	\Delta \mu^{\mathrm{TAS}}_{uv}(t) 
  = 
  \Tr\left\{\mathbf{P}_u \left[\mathbf{D}_v^{\mathrm{pp}}(t) - \mathbf{D}_v^{\mathrm{p}}(t)\right]\right\} \equiv \mu^{\mathrm{ind,pp}}_{uv}(t) - \mu^{\mathrm{ind,p}}_{uv}(t),
  \qquad
  u,v \in \{x,y,z\},
\end{equation}
where $\mu^{\mathrm{ind}}_{uv}(t)$ denotes the expectation value of the dipole
operator. The computation of TAS spectra involves performing two simulations
for recording the dipole moment at each time step; these simulations and their
quantities are denoted by p and pp subscripts, indicating that the real-time
propagation used pump-only pulse and pump together with the probe pulse,
respectively. This difference is calculated to isolate the effect of the
probe pulse on the non-stationary electronic wavepacket, \emph{i.e.} to
eliminate pump-only-dependent terms in Eq.~\eqref{eq:response0}. This
procedure is in contrast with the simulations initiated from an equilibrium
stationary state (\emph{i.e.} without pump), for which
Eq.~\eqref{eq:response0} simplifies to the static dipole moment that is
subtracted in Eq.~\eqref{eq:deltamu_TAS}, and the absorption spectra thus
only require one time propagation simulation.
The differential dipole moment is subsequently transformed into the frequency
domain (like in Eq.~\eqref{eq:nonpertResponse}).
Finally, the TAS spectral function $S^\mathrm{TAS}(\omega)$ is evaluated
analogously to its ground state absorption counterpart as,
\begin{align}
S^\mathrm{TAS}(\omega)
=
\frac{4\pi\omega}{3c} \Im\Tr[\boldsymbol{\alpha}^\mathrm{TAS}(\omega)]
~~&\mathrm{where,}~~
\alpha^\mathrm{TAS}_{uv}
=
\frac{\Delta \mu^{\mathrm{TAS}}_{uv}(\omega)}{\mathcal{F}_0}
~~.
\label{eq:SpectralS}
\end{align}
In Eq. \eqref{eq:SpectralS}, $c$ is the speed of light, $\Tr$ is a trace over
Cartesian components of the polarizability tensor $\boldsymbol{\alpha}$. 

In simulated X-ray spectra, signals originating from transitions between
valence and high-lying virtual orbitals can appear.  These are non-physical in
calculations using finite atom-centered basis sets, since above-ionization
states are ill-described in such cases.  To eliminate these spurious peaks, we
restrict the probe operator to act only on a selected subset of core and
virtual orbitals by zeroing all other elements of the dipole matrix when
applying the probe pulse thus making the non-physical transitions artificially
dipole-forbidden.  This technique,
called selective perturbation,
 was introduced in our previous works in the
context of XAS calculations using both RT-TDDFT~\cite{Kadek2015} as well as
damped linear response TDDFT~\cite{Konecny2022}.  Note that no restriction is
applied on the pump pulse as that is purposefully tuned to an excitation from
valence to low-lying orbitals.



In line with the spectral analysis technique for RT--TDDFT simulations
presented in Ref.~\citenum{Repisky2015}, we assign the nature of electronic
excitations underlying a particular TAS resonance feature by a dipole--weighted
transition matrix analysis (DWTA). For a resonant frequency $\omega^\prime$,
the Fourier component $\mathcal{T}(\omega^\prime)$ of the time--domain signal
$\mathcal{T}(t)$ contains all information. 
\begin{equation}
	\mathcal{T}(\omega^\prime) = \int_{0}^{\infty} \mathcal{T}(t+T+\tau) e^{(i\omega'-\Gamma)t} dt ~;
	~
	\mathcal{T}_{uv;ai}(t) = \Delta \mu_{uv;ai}^{\mathrm{TAS}}(t) 
	= \mu_{uv;ai}^\mathrm{ind,pp}(t) - \mu_{uv;ai}^\mathrm{ind,p}(t)
	~.
	\label{eq:anlaysis}
\end{equation}
Here, $a,i$ runs over occupied and virtual spinors, respectively, and we
accounted for the fact that only the time signal after the probe pulse is used
for the Fourier transform.  We use the differential induced dipole moment
(Eq~\eqref{eq:deltamu_TAS}) at a specific resonance frequency
($\omega^\prime$), to obtain the contribution of individual occupied--virtual
pairs towards the spectral peak. $\mathcal{T}(t)$ matrix is obtained by
averaging over the Cartesian components. 

The above formulated RT-TDDFT methodology was implemented in the \ReSpect ~program,~\cite{ReSpect} and used to investigate the 
TAS spectra of two prototypical molecules, ethylene and thiophene at the valence and core (L$_{2,3}$) energy region, respectively, employing 1c non-relativistic, 2c amfX2C and 4c Dirac--Coulomb Hamiltonian. Geometries of the systems are given in SI. 
Uncontracted aug-cc-pVXZ basis set (X=T(C), D(H) for ethylene and  T(S), D(C,H) for thiophene)~\cite{Dunning1989,Kendall1992,Woon1993,Wilson1999} with PBE0–40HF hybrid xc functional~\cite{Adamo1999} including 40\% exact exchange contributions is used.~\cite{Konecny2022}
For all nuclei, a finite-sized Gaussian model was used.
For core absorption processes, we use the selective perturbation scheme described above,
whereby only electric dipole moment contributions originating from the core S $2p_{1/2}$ and $2p_{3/2}$ orbitals are considered to be dipole-allowed.~\cite{Kadek2015}
The details of the TAS computational setup are elucidated in Table~\ref{tab:setup} and ~Figure~S1.
Note that the choice of direction of polarization of pump pulse ($\bm{n}$) is important as the spectra are sensitive to geometrical features. 

	\begin{table}[h]
	\begin{tabular}{ c| c c c c |c|cc}
		\hline
		\multirow{2}{*}{Molecule} & \multicolumn{4}{|c|}{$\bm{\mathcal{E}}(t)$} & $\bm{\mathcal{F}}(t)$ & \multicolumn{2}{|c}{RT simulation}\\
		& $\omega_0$ (au) & $T$ (au) & $\bm{n}$ & $\mathcal{E}_0 $ (au) & $\mathcal{F}_0 $ (au) & n$_\text{step}$ & $\Delta t_\text{step}$ (au)\\\hline
		C$_2$H$_4$ &  0.2762 & 341.10 & x & 0.01 & 0.01 & 20000 & 0.15\\
		C$_4$H$_4$S & 0.2135 & 441.20 & y & 0.05 & 0.01 & 25000 & 0.20\\
		\hline
	\end{tabular}
	\caption{Computational setup: The pump and probe pulse parameters corresponds to Eq. ~\eqref{eq:pump} and ~\eqref{eq:probe}. Carrier frequency ($\omega_0$) is tuned to the first bright excited state along the direction of dipole-allowed transition ($\bm{n}$). Time duration ($T$) of the pump pulse corresponds to 15 carrier periods. The pump pulse amplitude ($\mathcal{E}_0 $) is chosen to sufficiently depopulate the ground state. A weak $\delta$ probe of amplitude $\mathcal{F}_0 $ along all the Cartesian components are applied. 
The probed system is evolved for n$_\text{step}$ with $\Delta t_\text{step}$ length
as detailed in Section S1, using the convergence criterion for micro-iterations
$|\mathbf{D}^{(n)}(t+\Delta t) - \mathbf{D}^{(n+1)}(t+\Delta t)| < 10^{-6}$.
These parameters are used unless otherwise stated. 
 }
	\label{tab:setup}
\end{table}
\begin{figure}[htb!]
	\centering
	\includegraphics[width=0.4\textwidth]{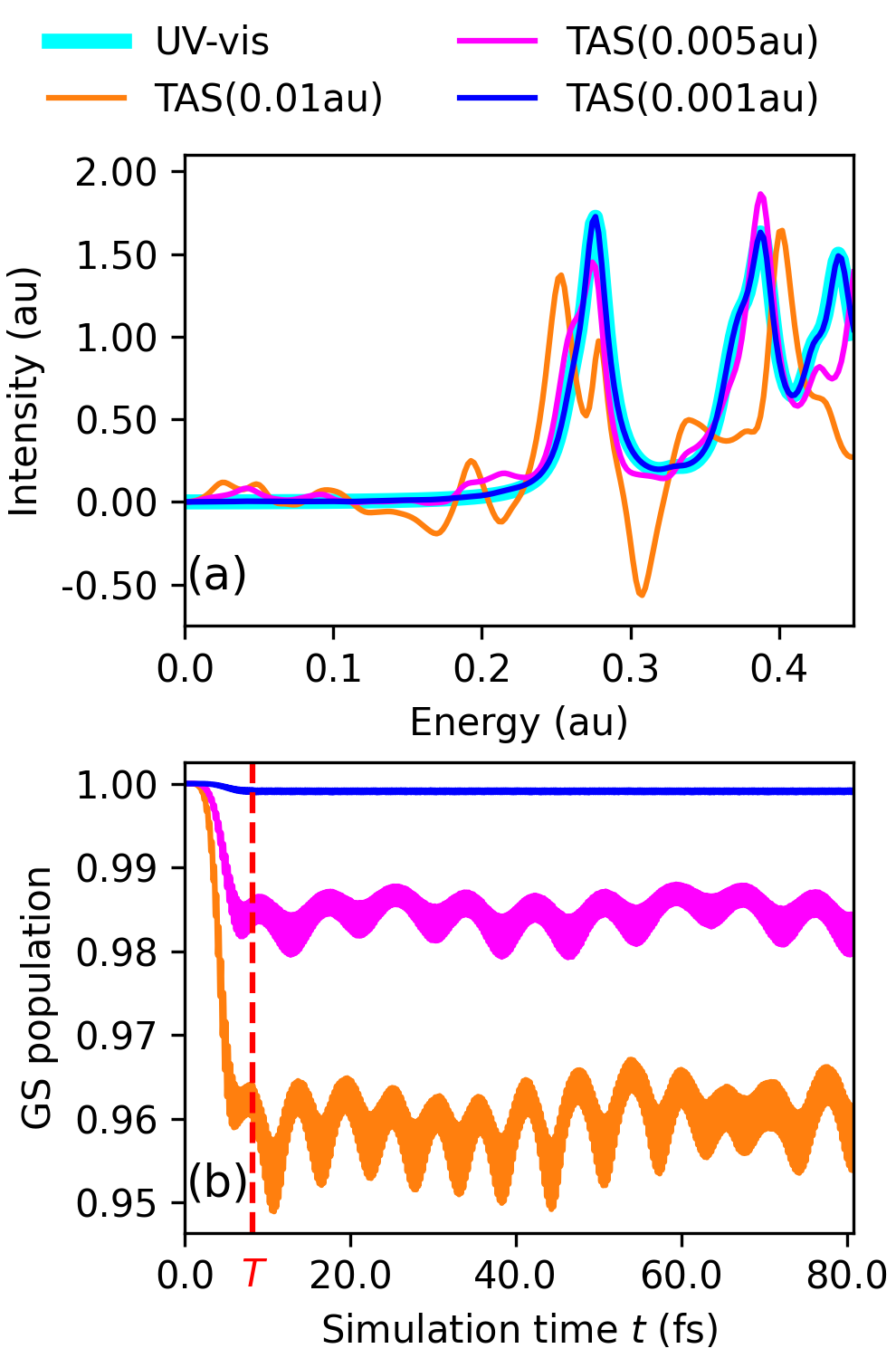}
	\caption{Ethylene -- (a) Comparison of UV-vis and TAS spectra with varying $\mathcal{E}_0 $, given in parentheses, obtained with amfX2C Hamiltonian. 
TAS is computed at $\tau=0.0$ with a damping parameter $\Gamma = 0.01$ au. 
		(b) Variation in the ground state population as a function of simulation time $t$ 
		obtained as $\Tr[\mathbf{D}(0) \mathbf{D}(t)]$,
		with pump amplitudes as color coded  in (a).
		$T$ (in red) marks the duration of pump pulse.
	}
	\label{fig:ethylene-pop}
\end{figure}
For ethylene, we study the effect of the degrees of freedom specific to pump-probe processes, focusing on pump pulse strength and pump-probe time delay.
Firstly, the effect of pump pulse strength on the TAS spectra is shown in Figure~\ref{fig:ethylene-pop}. 
The electronic wavepacket $\ket{\Psi[\bm{\mathcal{E}}]}$ generated by the pump pulse comprises of an admixture of ground and excited electronic states (see Eq.~\eqref{eq:wavepacket}), with the effective depopulation of ground state  being proportional to $\mathcal{E}_0$.
The ground state occupancy at time $t$ is estimated as $\Tr[\mathbf{D}(0) \mathbf{D}(t)]$.~\cite{Walkenhorst2016}
It is evident from Figure~\ref{fig:ethylene-pop}, that considerable depopulation of ground state is necessary to differentiate TAS spectra from ordinary ground state absorption spectra. 
The negative intensity signal at about 0.3 au and $\mathcal{E}_0=0.01$ au in Figure~\ref{fig:ethylene-pop}a is a hallmark of non-stationary state involved in this pump-probe process and is a consequence of mixing Lorentzian and Rayleigh lineshape functions as derived in Eq.~\eqref{eq:response1Imag}.
TAS spectra of ethylene obtained with $\mathcal{E}_0=0.01$ au is in a good agreement with the pioneering work by \citeauthor{DeGiovannini2013}~\cite{DeGiovannini2013} using Octopus code~\cite{marques2003octopus,castro2006octopus,tancogne2020octopus}.
However, due to different computational setup and lack of experimental references we do not focus on a detailed assignment of the TAS spectral features. 
For exact theory, the ground state population at $t> T+\tau$ depends only on $\tau$  and pulse intensities, and is constant with respect to the simulation time $t$.
This can be derived from Eqs.~\eqref{eq:evolWaveFunc}~and~\eqref{eq:wavepacket} as,
\begin{align}
	| c_0(\bm{\mathcal{E}}) |^2 &= |\bra{\Phi_0}\ket{\Psi(t)} |^2 \nonumber \\
	&= |\bra{\Phi_0} e^{-i\hat{H}_0(t-T-\tau)} e^{-i\hat{Q}_v} e^{-i\hat{H}_0\tau} \ket{\Psi[\bm{\mathcal{E}}]} |^2 \nonumber \\
	&=  |\bra{\Phi_0} e^{-i\hat{Q}_v} e^{-i\hat{H}_0\tau} \ket{\Psi[\bm{\mathcal{E}}]} |^2  ~.
	\label{eq:gs-pop-exacttheory}
\end{align}
However, this behaviour does not hold for approximate mean-field methods such as HF or DFT,
where the exponential term $e^{-i\hat{H}_0(t-T-\tau)}$ does not drop out of the integration as a consequence of the implicit time dependence incorporated in the Fock matrix. The resulting oscillations in the ground state population,  as seen in Figure~\ref{fig:ethylene-pop}b, are artefacts attributed to the use of adiabatic approximation.~\cite{PhysRevB.84.075107,PhysRevLett.114.183002,C8CP03957G}

\begin{figure}
	\centering
	\includegraphics[width=0.95\textwidth]{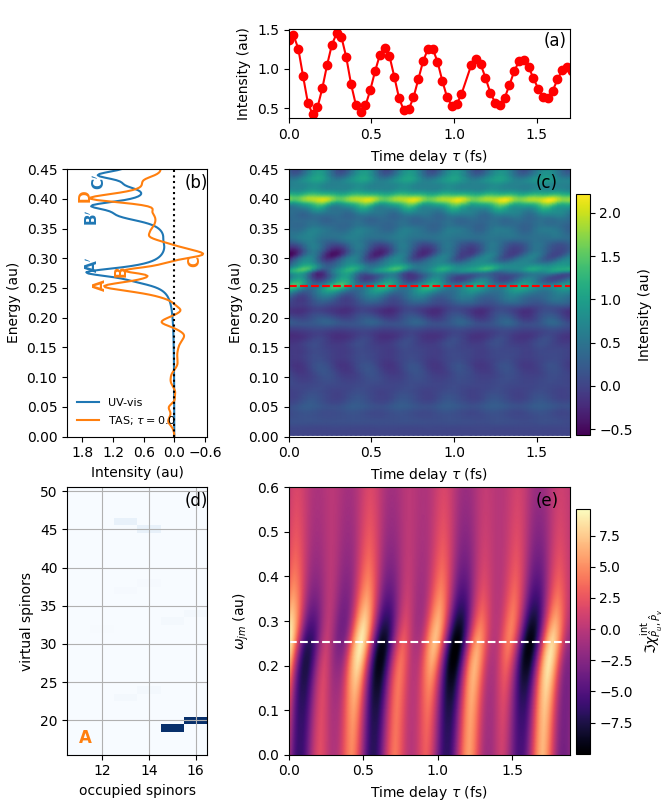}
	\caption{Ethylene --
		(a) Variation in intensity with $\tau$ for the TAS spectral peak at $\omega_0 = 0.2534$ au;
		(b) Ground state absorption and TAS spectra at $\tau= 0.0$;
		(c) TAS spectra with varying $\tau$;
		(d) DWTA of the TAS peak \textbf{A} at $\omega_0 = 0.2534$ au and $\tau =0.0$. The intensity of blue color corresponds to the intensity of the particular excitation;
		(e) $ \Im\chi^\mathrm{int}_{\hat{P}_u,\hat{P}_v} [\bm{\mathcal{E}}] (0.2534,341.10+\tau) $ is  given in Eq. \eqref{eq:F-plot},
		for degenerate $k,j$ spinor-pair selected as (15,16)$\to$(19,20) and (13,14)$\to$(46,45)
		with orbital energy difference of 0.3305 au and 0.6039 au, respectively. White dashed lines corresponds to the resonant $\omega$. 2c amfX2C Hamiltonian is used for the RT simulation.
	}
	\label{fig:ethylene-heatmap}
\end{figure}

From an experimental point-of-view, the most important degree of freedom in TAS is the pump--probe time delay $\tau$.
The influence of $\tau$ on TAS for ethylene is depicted in Figure~\ref{fig:ethylene-heatmap}c.
As the pump pulse is tuned to the first bright state in the ground state absorption spectrum (\textbf{A}$^\prime$ in Figure~\ref{fig:ethylene-heatmap}b), 
this ground state peak is split into peaks \textbf{A}, \textbf{B} and \textbf{C} in the TAS. 
DWTA analysis of \textbf{A}$^\prime$, \textbf{A}, \textbf{B} and \textbf{C} (shown in Figure~\ref{fig:ethylene-heatmap}d,~S3~and~S4) shares same dominantly contributing transition of $\pi\to\pi^\ast$ character involving degenerate spinor pair (15,16) $\to$ (19,20). Secondary contributions come from transitions involving degenerate spinor pair (13,14)$\to$(46,45).
Interestingly, peak \textbf{D} has completely different character from any of the features obtained in the ground state absorption spectra, further emphasising on the novelty of this spectroscopic technique to probe states (and therein molecular orbitals) not directly accessible by purely ground or excited-state absorption. 

A striking feature of the TAS spectrum in Figure~\ref{fig:ethylene-heatmap} is the oscillation of spectral intensity with $\tau$.
As shown in Figure~S5, this feature is not associated with the variation of
the contributing orbitals underlying a particular frequency peak with time-delay.
Therefore, non-equilibrium response theory was applied to understand qualitatively the origin of these modulations.
Considering that the system consists of only light elements (C,H), a reasonable approach is to use only the non-relativistic component associated with the real part of the response function in Eq.~\eqref{eq:response1Imag}.
As shown in Figure~\ref{fig:ethylene-heatmap}d, the DWTA of peak \textbf{A} reveals that
only two transitions $[(15,16) \to(19,20)]$ and $[(13,14)\to(46,45)]$ with orbital energy difference of 0.3305 au and 0.6039 au contribute. Therefore, only these transitions appear in the summation over $k,j$ in Eq.~\eqref{eq:response1Imag}.
By discretizing the summation over $m$ such that $\omega_{jm}$ is within the frequency range [0.0,0.6au], and further assuming that the two transitions:
(I) are of equal probability, i.e. $|c_j(\bm{\mathcal{E}})c_k(\bm{\mathcal{E}})| = 1$;
(II) have zero phase-factor difference ($\phi_j - \phi_k =0$);
(III) have the same electric dipole transition moments for all $m$ ($P_{v,km}=1$),
we arrive at a simplified formula for the response function, plotted in Figure~\ref{fig:ethylene-heatmap}e,
\begin{align}
	\Im\chi^\mathrm{int}_{\hat{P}_u,\hat{P}_v} [\bm{\mathcal{E}}] (0.2534,341.10+\tau) 
  &=\sum_{\omega_{kj}\in 0.3305,0.6039} 
  ~
  \cos(-\omega_{kj}\tau)
  \Big\{ \mathcal{L}(0.2534 - \omega_{jm}) - \mathcal{L}(0.2534 + \omega_{jm}) \Big\}
  \nonumber 
  \\
  &+ 
  \sin(-\omega_{kj}\tau) 
  \Big\{ \mathcal{R}(0.2534 - \omega_{jm}) + \mathcal{R}(0.2534 + \omega_{jm}) \Big\}
  ~.
  \label{eq:F-plot}
\end{align}
Even with such simplifications, 
we were able to mimic the appearance of the intensity oscillation with $\tau$,
unique to TAS.
Especially we are capable of capturing the decrease in intensity at the maxima at $\omega=0.2534$ au, marked by white dashed line in Figure~\ref{fig:ethylene-heatmap}e.
Providing more 
 information  about initial phase difference and dipole contributions, will enable one to more closely reproduce the spectral feature from response theory.

\begin{figure}[htb!]
	\centering
	\includegraphics[width=0.5\textwidth]{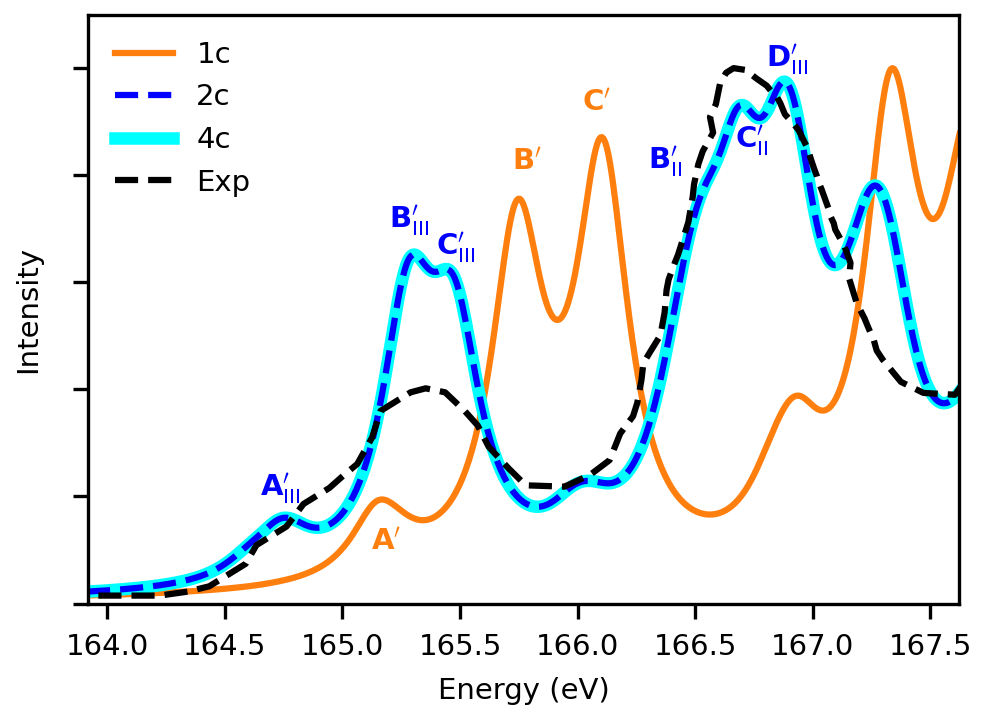}
	\caption{Thiophene -- Ground state L$_{2,3}$-edge X-ray absorption spectra obtained with 1c non-relativistic (\emph{orange}), 2c amfX2C (\emph{blue}) and  4c Dirac--Coulomb (\emph{cyan}) Hamiltonian using a damping factor of $\Gamma = 0.005$ au. 
		The simulated spectra is shifted by +1.42 eV, to match the first experimental peak. 
		Experimental results are digitized from Ref.~\citenum{Baseggio2017-exp}.
		DWTA analysis of the spectral peaks obtained with 1c and 2c Hamiltonians are shown in Figures S6 and S7, respectively.
	}
	\label{fig:thiophene-method}
\end{figure}
\begin{figure}[htb!]
	\centering
	\includegraphics[width=0.95\textwidth]{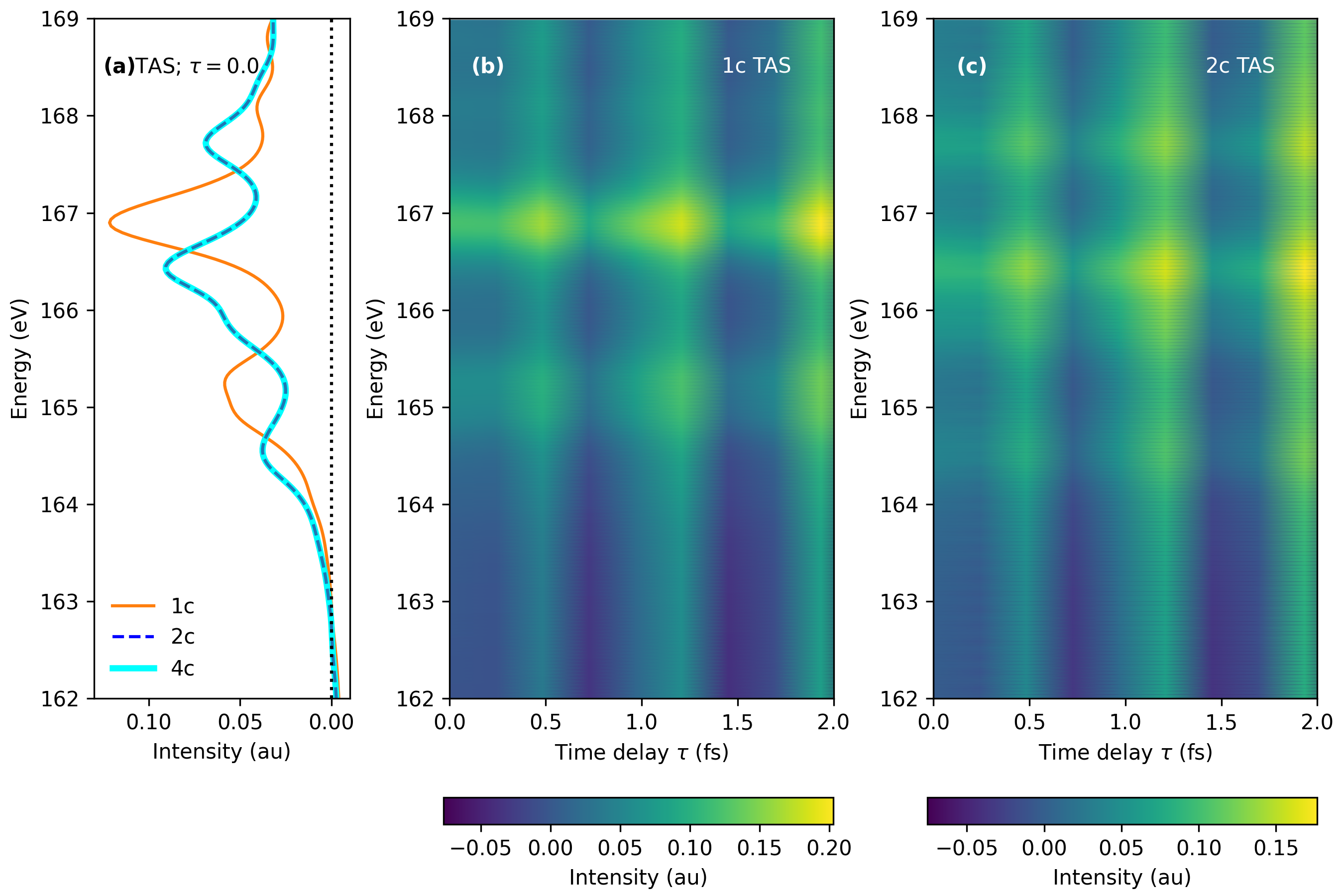}
	\caption{Thiophene -- 
		(a) TAS at $\tau=0$ with 1c non-relativistic (\emph{orange}), 2c amfX2C (\emph{blue}) and 4c Dirac--Coulomb (\emph{cyan}) Hamiltonian;
		(b) Variation in TAS spectra with $\tau$ obtained with 1c non-relativistic Hamiltonian;
		(c) Variation in TAS spectra with $\tau$ obtained with 2c amfX2C Hamiltonian.
		All spectra are obtained with a damping factor $\Gamma = 0.01$ au.
	}
	\label{fig:thiophene-hm}
\end{figure}
Next, we focus on the energy region of the core L$_{2,3}$-edge of sulphur in thiophene.
The experimental~\cite{Baseggio2017-exp} and simulated ground state XAS at the non-relativistic (1c), amfX2C (2c), and Dirac--Coulomb (4c) Hamiltoanian level are presented in Figure~\ref{fig:thiophene-method},
clearly demonstrating the importance of including SO coupling effects to reproduce the doublet structure of the spectra (\textbf{B$^\prime_{\mathrm{III}}$}, \textbf{C$^\prime_{\mathrm{III}}$} versus 
	\textbf{B$^\prime_{\mathrm{II}}$}, \textbf{C$^\prime_{\mathrm{II}}$}
	). 
The DWTA analysis of individual peaks (shown in Figure~S6 and~S7)
 reveals that the lower energy feature corresponds to promotion from the S $2p_{3/2}$, whereas the higher energy ones are from S $2p_{1/2}$. 
Each of the SO split band, is further split due to the molecular field (MF), corresponding to $9e_{1/2}$, $8e_{1/2}$ (L$_{3}$ component) and $7e_{1/2}$ (L$_{2}$ component) molecular orbitals and leading to the lower energy separated doublets.
The MF doublet (\textbf{B}$^\prime$ and \textbf{C}$^\prime$) also appears at the 1c level.
 The SO split counterpart ~\textbf{A$^\prime_{\mathrm{II}}$} of band~\textbf{A$^\prime_{\mathrm{III}}$} is hidden under the more intense \textbf{B$^\prime_{\mathrm{III}}$} and \textbf{C$^\prime_{\mathrm{III}}$} doublet. 
 The experimental SO splitting of 1.2 eV is well reproduced by our relativistic Hamiltonian.
  We would like to draw the attention of the readers to the remarkable agreement between the XAS obtained with our amfX2C and gold-standard Dirac--Coulomb Hamiltonian, the former being generated at less than $1/7$th of the computational cost of the later.
 In addition, all other computational characteristics of RT simulations such as number of micro-iterations and convergence pattern remain identical for 2c and 4c regime.
Finally, we compute the TAS spectra using a setup described in Table~\ref{tab:setup} and shown in Figure~\ref{fig:thiophene-hm}.
The importance of incorporating relativistic effects is further reinforced here as evident from the SO split doublet spectral peaks generated using the relativistic Hamiltonian, as seen by comparing Figure~\ref{fig:thiophene-hm}b~and~\ref{fig:thiophene-hm}c.
Again, the resemblance of the TAS simulated at the 2c and 4c level at $\tau=0.0$ (in Figure~\ref{fig:thiophene-hm}a) further gives us confidence to use the modern amfX2C Hamiltonian for larger systems in future studies.


In summary, we have formulated and implemented a theoretical approach for
first-principle simulation of TAS spectra based on relativistic RT-TDDFT
frameworks, consistently applicable across the Periodic Table and
element-specific core energy regions.  Alongside gold-standard four-component
methodology, remarkable accuracy and significant computational acceleration was
achieved by introducing amfX2C Hamiltonian within RT-TDDFT framework.  With
this, we identify and interpret the unique features associated with TAS
spectra, notably the appearance of negative intensity peaks and oscillations in
intensity of a particular energy feature with pump-probe time delay. These
observations were further supported by non-equilibrium response theory,
relativistic generalization of which was also formulated in this letter.
Finally, we showcase fingerprints of spin-orbit effects on the X(T)AS
spectra near the sulphur L$_{2,3}$-edge of thiophene.  We believe, this work
constitutes a significant methodological advancement for studying and
interpreting transient absorption spectra, applicable to X-ray regimes and/or
heavy systems. Work along this direction is currently ongoing in our laboratory.


\begin{acknowledgement}
We acknowledge the support received from the Research Council of
Norway through a Centre of Excellence Grant (No.~262695), a Research Grant 
(No.~315822), and Mobility Grants (No.~301864 and No.~314814), 
as well as the use of computational resources provided by UNINETT Sigma2 -- The National
Infrastructure for High Performance Computing and Data Storage
in Norway (Grant No. NN4654K). In addition, this project received
funding from the European Union’s Horizon 2020 research and innovation
program under the Marie Skłodowska-Curie Grant Agreement No. 945478 (SASPRO2),
and the Slovak Research and Development Agency (Grant No. APVV-21-0497).
\end{acknowledgement}


\begin{suppinfo}
Real--time propagator;
Computational setup; 
Dependence of TAS on pump pulse carrier frequency; 
DWTA;
Molecular geometries.
This information is available free of charge via the Internet at http://pubs.acs.org/.

\end{suppinfo}

\bibliography{bibliography}


\end{document}